\documentclass[aps, preprint, groupedaddress, floatfix]{revtex4-1}

\usepackage[utf8]{inputenc}
\usepackage{epsfig}
\usepackage{epstopdf}
\usepackage{graphicx}
\usepackage{amsmath}
\usepackage{color}
\usepackage[export]{adjustbox}
\usepackage[colorlinks=true, allcolors=blue]{hyperref}
\usepackage{lineno}
\begin{document}

\title{Ferromagnetic polar metals via epitaxial strain: a case study of SrCoO$_3$}

\author{Zhiwei Liu$^{1,2,3}$, Qiuyue Li$^{1,2,3}$ and Hanghui Chen$^{2,3,4,*}$}

\affiliation{\mbox{$^1$Key Laboratory of Polar Materials and Devices, Ministry of Education,} \mbox{East China Normal University, Shanghai 200241, China}\\
  \mbox{$^2$NYU-ECNU Institute of Physics, NYU Shanghai, Shanghai 200124, China}\\
  \mbox{$^3$Department of Electronic Science, East China Normal University, Shanghai 200241, China}\\
  \mbox{$^4$Department of Physics, New York University, New York 10012, USA}\\
  $^*$Email: hanghui.chen@nyu.edu
}

\date{\today}

\begin{abstract}
  While polar metals are a metallic analogue of ferroelectrics, magnetic polar metals can be considered as a metallic analogue of multiferroics. There have been a number of attempts to integrate magnetism into a polar metal by synthesizing new materials or heterostructures. Here we use a simple yet widely used approach--epitaxial strain in the search for intrinsic magnetic polar metals. Via first-principles calculations, we study strain engineering of a ferromagnetic metallic oxide SrCoO$_3$, whose bulk form crystallizes in a cubic structure. We find that under an experimentally feasible biaxial strain on the $ab$ plane, collective Co polar displacements are stabilized in SrCoO$_3$. Specifically, a compressive strain stabilizes Co polar displacements along the $c$ axis, while a tensile strain stabilizes Co polar displacements along the diagonal line in the $ab$ plane. In both cases, we find an intrinsic ferromagnetic polar metallic state in SrCoO$_3$. In addition, we also find that a sufficiently large biaxial strain ($> 4\%$) can yield a ferromagnetic-to-antiferromagnetic transition in SrCoO$_3$. Our work demonstrates that in addition to yielding emergent multiferroics, epitaxial strain is also a viable approach to inducing magnetic polar metallic states in quantum materials.
\end{abstract}

\maketitle

\section{Introduction}

Ferroelectric~\cite{10.1038/358136a0,10.1103/PhysRevLett.92.257201,10.1103/PhysRevB.53.1193,10.1021/ja0758436} and multiferroic~\cite{10.1126/science.1080615,10.1038/nmat1804,10.1038/nmat1805,10.1126/science.1094207} materials have wide applications. By definition, they are insulators with a spontaneous polarization below a Curie temperature and the polarization is switchable by an external electric field~\cite{10.1126/science.1159655}. Usually, it is difficult to stabilize a macroscopic polarization in metals since itinerant electrons screen internal dipoles and suppress spontaneous polar displacements. In 1965, Anderson and Blount proposed that if the optical phonons that are responsible for the polar displacements have weak couplings to itinerant electrons, a ferroelectric-like structural phase transition may occur even in a metal at finite temperatures~\cite{10.1103/PhysRevLett.14.217}. The prediction was confirmed in 2013 when Shi et al. successfully synthesized LiOsO$_3$ and observed a continuous centrosymmetric-to-polar structural phase transition around 140 K~\cite{10.1038/nmat3754}. Since then, the study of polar metals, the metallic analogue of ferroelectrics, has drawn great attention~\cite{10.1021/acs.nanolett.8b00633,10.1038/nature17628,10.1038/s41598-017-04635-3,10.1038/s43246-019-0005-6,10.1063/1.5035133,10.1103/PhysRevMaterials.3.054405}, not only because of interests in basic sciences~\cite{10.1038/s41467-020-18438-0} but also in potential technological applications such as electrodes in ferroelectric nanocapacitors~\cite{10.1063/1.5049607}. More recently, a number of attempts have been made to integrate magnetism into a polar metal~\cite{10.1103/PhysRevB.102.144418,10.1103/PhysRevB.99.224411,10.1038/s41467-019-13270-7,10.1103/PhysRevMaterials.6.044403,10.1126/sciadv.abm7103}. A magnetic polar metal is the metallic analogue of multiferroics. Experimentally, magnetic polar metallic states have been found in Pb$_2$CoOsO$_6$~\cite{10.1103/PhysRevB.102.144418}, Fe-doped Ca$_3$Ru$_2$O$_7$~\cite{10.1103/PhysRevB.99.224411}, BaTiO$_3$/SrRuO$_3$/BaTiO$_3$ heterostructure~\cite{10.1038/s41467-019-13270-7} and $AA'$-stacked (Fe$_{0.5}$Co$_{0.5}$)$_5$GeTe$_2$~\cite{10.1103/PhysRevMaterials.6.044403,10.1126/sciadv.abm7103}. However, all these materials either have complicated chemical composition or are artificial heterostructures. On the other hand, epitaxial strain has been widely used to tune physical properties of quantum materials~\cite{10.1073/pnas.2101946118}, in particular polarization~\cite{10.1146/annurev.matsci.37.061206.113016}. Well-known examples include strain-induced ferroelectricity in SrTiO$_3$, EuTiO$_3$ and SrMnO$_3$~\cite{10.1038/nature09331,10.1103/PhysRevLett.97.267602,10.1103/PhysRevLett.104.207204,10.1038/nature02773}, as well as a strain-driven morphotropic phase boundary in BiFeO$_3$~\cite{10.1126/science.1177046}. Thus, it is worthwhile to study whether strain engineering can also be utilized to induce a magnetic polar metallic state in known materials.
 
In this work, we demonstrate how to use epitaxial strain to stabilize an intrinsic ferromagnetic polar metallic state in a known complex oxide SrCoO$_3$. Bulk SrCoO$_3$ is a metal and crystallizes in a simple cubic structure~\cite{10.1103/PhysRevB.51.11501}. It exhibits ferromagnetic order below $T_c = 305$ K~\cite{10.1088/0953-8984/23/24/245601}. We use first-principles calculations and find that an experimentally feasible biaxial strain, either compressive or tensile, can break inversion symmetry in metallic SrCoO$_3$. Specifically, we find that under a biaxial compressive strain of 2.4\% to 4\% imposed on the $ab$ plane, collective Co displacements along the $c$ axis are stabilized, while under a biaxial tensile strain of 2.9\% to 4\% imposed on the $ab$ plane, collective Co displacements along the diagonal line in the $ab$ plane are stabilized. In both cases, inversion symmetry is broken via a centrosymmetric-to-polar structural transition and an intrinsic ferromagnetic polar metallic state is induced in SrCoO$_3$. Furthermore, we find that under a sufficiently large biaxial strain ($> 4\%$), a ferromagnetic-to-antiferromagnetic transition may occur to SrCoO$_3$. Our work shows that in addition to synthesizing new materials or heterostructures, we can also use strain engineering as a viable approach to searching for magnetic polar metals.

\section{Computational details}
We perform density functional theory
(DFT)~\cite{10.1103/PhysRev.136.B864,10.1103/PhysRev.140.A1133}
calculations, as implemented in Vienna Ab Initio Simulation Package
(VASP)~\cite{10.1103/PhysRevB.49.14251,10.1103/PhysRevB.54.11169}. We
use the generalized gradient approximation with the
Perdew-Burke-Ernzerhof parameterization
(GGA-PBE)~\cite{10.1103/PhysRevLett.77.3865} as the
exchange-correlation functional. An energy cutoff of 600 eV is used
throughout the calculations. The Brillouin
  zone~\cite{10.1103/PhysRevB.13.5188} integration is performed with a
  Gaussian smearing of 0.05 eV over a Monkhorst-Pack \textbf{k}-mesh
  of $16\times16\times16$ for 5-atom simulation cell and a
  Monkhorst-Pack \textbf{k}-mesh of $10\times10\times8$ for
  $\sqrt{2}\times\sqrt{2}\times2$ 20-atom supercell. The convergence
threshold for the self-consistent calculation is $10^{-8}$ eV. Atomic
relaxation is converged when each force component is smaller than
$10^{-3}$ eV/\AA~and pressure on the simulation cell is less than 0.5
kbar. We use the finite-displacement
method~\cite{10.1209/0295-5075/32/9/005,10.1103/PhysRevLett.78.4063}
with the aid of Phonopy~\cite{10.1016/j.scriptamat.2015.07.021} to
calculate phonon band structure and phonon density of states.  For
biaxial strain calculations, we fix the two in-plane lattice constants
($a$ and $b$) and allow the out-of-plane lattice constant ($c$ axis)
to fully relax. The strain is defined as $\xi
=(a-a_{\textrm{opt}}$)/$a_{\textrm{opt}}$$\times$100\% where
$a_{\textrm{opt}}$ is the DFT optimized lattice constant of
ferromagnetic cubic SrCoO$_3$ and $a$ is the theoretical lattice
constant of the substrate that imposes biaxial
strain. All the calculations are spin polarized. To
  consider the correlation effects in SrCoO$_3$, we test a range of
  $U$ on Co-$d$ orbitals using the spin polarized DFT+$U$ method. We
  find that for both lattice constant and magnetization, $U=0$ yields
  the best agreement between theory and experiment. Increasing $U$,
  however, impairs the agreement. Therefore we choose $U=0$ in our
  calculations. See Supplemental Material~\cite{SI} Note 1 for details. Since polarity
and magnetism both originate from Co atom and the atomic number of Co
is small, we neglect spin-orbit interaction in this study. We use the Aflow library (Automatic FLOW for Materials Discovery \href{https://www.aflowlib.org/aflow-online}{www.aflowlib.org/aflow-online}) to determine the space group of various SrCoO$_3$ crystal structures and also use Phononpy to cross-check it.

\section{Results}

\begin{figure}[t]
    \centering
    \includegraphics[width=0.8\textwidth]{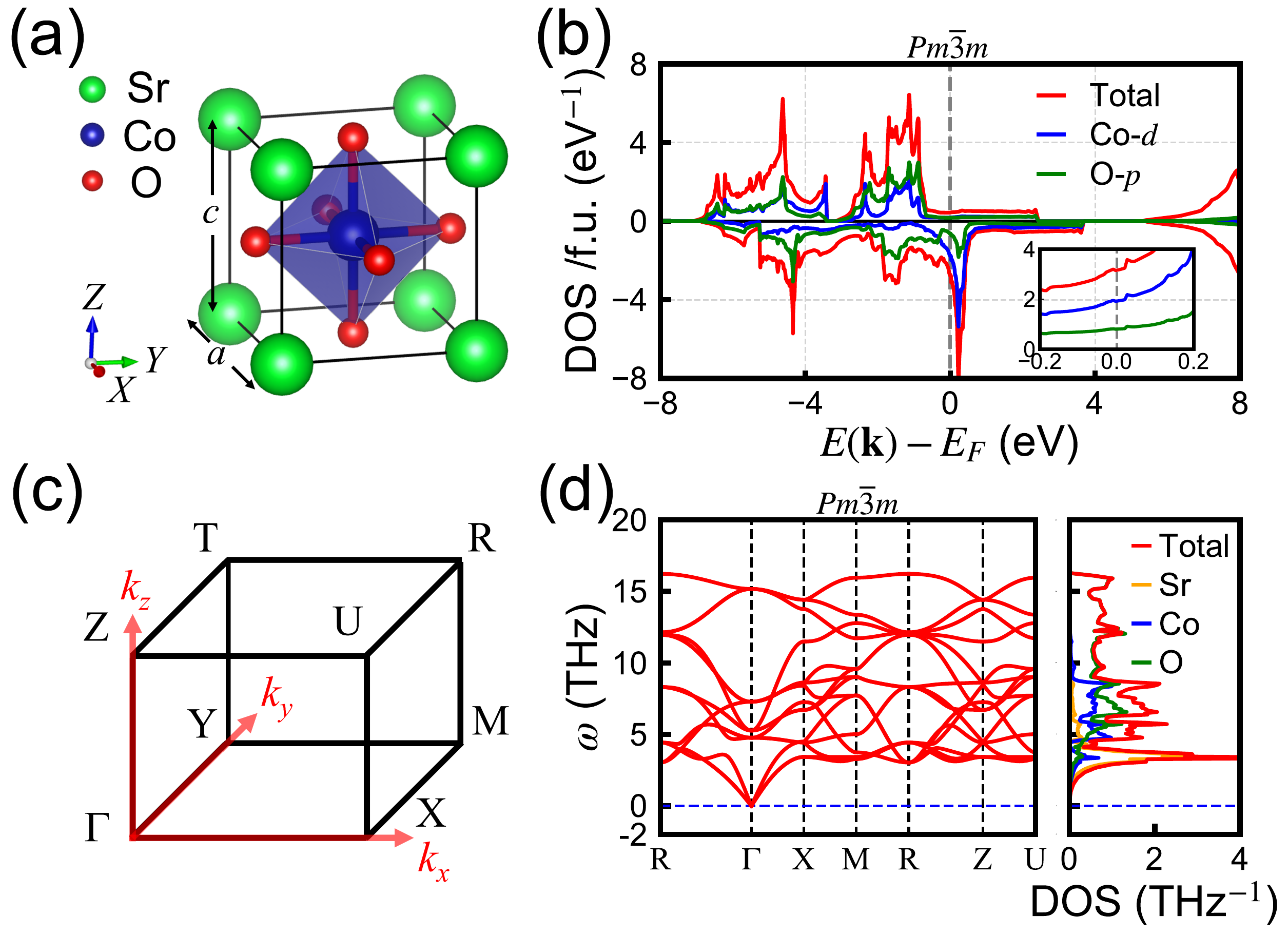}
    \caption{(a): The crystal structure of cubic SrCoO$_3$. The green, blue and red balls represent Sr, Co and O atoms, respectively.  (b): Density of states (DOS) of cubic SrCoO$_3$. The red, blue and green curves are total, Co-$d$ projected and O-$p$ projected DOS. Inset: near-Fermi-level DOS (spin up + spin down) of cubic SrCoO$_3$. (c): $\frac{1}{8}$ Brillouin zone of cubic SrCoO$_3$. (d): Phonon band structure and phonon DOS of cubic SrCoO$_3$. The coordinates of the high-symmetry \textbf{k}-points are: $R$(0.5, 0.5, 0.5), $\Gamma$(0.0, 0.0, 0.0), $X$(0.5, 0.0, 0.0), $M$(0.5, 0.5, 0.0), $Z$(0.0, 0.0, 0.5), $U$(0.5, 0.0, 0.5). The red, orange, blue and green curves are total, Sr projected, Co projected and O projected phonon DOS.}
    \label{fig1}
\end{figure}

\subsection{Bulk}
First, we calculate bulk properties of SrCoO$_3$. Experimentally, SrCoO$_3$ crystalizes in a simple cubic structure without any oxygen octahedral rotations. The corresponding space group is $Pm\bar{3}m$ (no. 221) and the corresponding Glazer notation is $a^0a^0a^0$~\cite{10.1107/S0108768102015756}. Bulk SrCoO$_3$ is a ferromagnetic metal below 305 K with a saturation magnetic moment of 2.5 $\mu_{\mathrm{B}}$/f.u. at 2 K~\cite{10.1088/0953-8984/23/24/245601}. Fig.~\ref{fig1}(a) shows the optimized crystal structure of bulk SrCoO$_3$ in our DFT calculations. We find that it is stabilized in a cubic structure with the optimized lattice constant $a_{\textrm{opt}} = 3.827$~\AA, which is in good agreement with experiment and the previous theoretical studies~\cite{10.1002/zaac.19936190104,10.1088/0953-8984/23/24/245601,10.1103/PhysRevB.57.13655,10.1103/PhysRevB.60.16423,10.1103/PhysRevX.10.021030}. Fig.~\ref{fig1}(b) shows the density of states (DOS) of bulk SrCoO$_3$. The DOS clearly shows a ferromagnetic metallic state with the exchange splitting being about 1 eV. Around the Fermi level, there are Co-$d$ and O-$p$ states, which are strongly hybridized with each other. Fig.~\ref{fig1}(c) shows the Brillouin zone of an orthogonal crystal structure (cubic structure is a special case), in which all the high-symmetry \textbf{k}-points are labelled. They are used in this figure as well as in the subsequent figures. Fig.~\ref{fig1}(d) shows the phonon spectrum and density of states of cubic SrCoO$_3$. We find that cubic SrCoO$_3$ is free from imaginary phonon modes, indicating that the cubic structure is stable. From the phonon spectrum, we find that the low-frequency phonons are mainly associated with the vibration of Sr atoms since Sr atoms are the heaviest among SrCoO$_3$, while the high-frequency phonons are associated with the vibration of O atoms because O atoms have light mass. The above results show that DFT provides a reasonable description of electronic and structural properties of bulk SrCoO$_3$.

%the cubic SrCoO$_3$ ($\cal{SG}$ No. 221 $Pm\bar{3}m$, Glazer notation $a^0a^0a^0$), which without any oxygen octahedral rotation (shown in panel \textbf{a} of Fig. 1). In our calculations, after fully relaxation, the lattice constant of cubic SrCoO$_3$ is 3.827 \AA, and \textcolor{blue}{the total moment} is 2.6 $\mu_{\mathrm{B}}$/f.u., which are good agreement with the previous work~\cite{10.1002/zaac.19936190104,10.1088/0953-8984/23/24/245601,10.1103/PhysRevB.57.13655,10.1103/PhysRevB.60.16423}. Panel \textbf{b} of Fig. 1 shows the density of states of the cubic SrCoO$_3$. We can see that the Co-$d$ and O-$p$ orbitals have states cross the Fermi level, this indicates that cubic SrCoO$_3$ exhibits metallic properties. This is consistent with the previous experimental results~\cite{10.1088/0953-8984/23/24/245601}. Panel \textbf{c} of Fig. 1 shows $\frac{1}{8}$ Brillouin zone of the cubic SrCoO$_3$. And we calculate the phonon band structure and density of states of cubic SrCoO$_3$ (shown in panel \textbf{d}). We find there are no negative phonon frequency. This result indicates the stability of cubic SrCoO$_3$.

\subsection{Compressive strain}

\begin{figure}[h]
    \centering
    \includegraphics[width=0.8\textwidth]{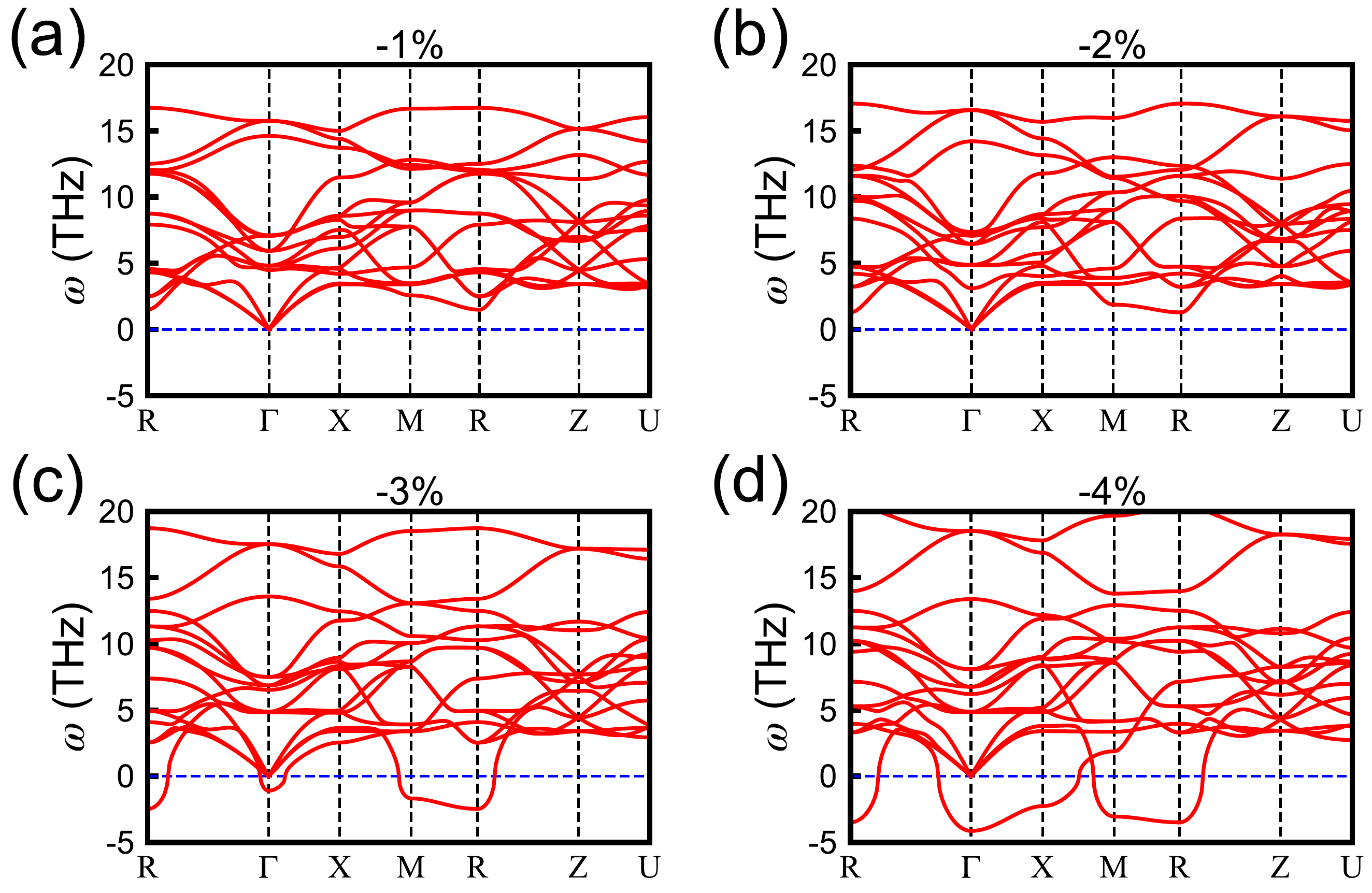}
    \caption{Phonon band structure of SrCoO$_3$ under different compressive strains. (a): 1$\%$ compressive strain.
    (b): 2$\%$ compressive strain. (c): 3$\%$ compressive strain. (d): 4$\%$ compressive strain. The coordinates of the high-symmetry \textbf{k}-points are: $R$(0.5, 0.5, 0.5), $\Gamma$(0.0, 0.0, 0.0), $X$(0.5, 0.0, 0.0), $M$(0.5, 0.5, 0.0), $Z$(0.0, 0.0, 0.5), $U$(0.5, 0.0, 0.5).}
    \label{fig2}
\end{figure}

Next we study SrCoO$_3$ under compressive biaxial strain. Under a compressive strain, cubic SrCoO$_3$ naturally transforms to a tetragonal structure with $c/a > 1$. However, the simple tetragonal structure with no other distortions is not necessarily dynamically stable. To carefully check this point, we perform phonon calculations on the simple tetragonal structure of SrCoO$_3$ under various compressive strains. We show the results in Fig.~\ref{fig2}. We find that under 1\% and 2\% compressive strains, SrCoO$_3$ is stabilized in a simple tetragonal structure with no other distortions. This corresponds to space group $P4/mmm$ (no. 123) and Glazer notation ($a^0a^0c^0$). However, under 3\% and 4\% compressive strains, imaginary phonon modes appear at $\Gamma$, $M$, $R$ and $X$ points, indicating that other structural distortions may occur in the simple tetragonal structure~\cite{10.1088/2516-1075/ac78b3}. By analyzing the vibration modes, we find that the imaginary phonon at $\Gamma$ point is a polar mode with the Co and O atoms moving out-of-phase along the $c$-axis; the imaginary phonon at $M$ point is a mode in which the CoO$_6$ oxygen octahedra rotate in-phase about the $c$-axis; the imaginary phonon at $R$ point is a mode in which the CoO$_6$ oxygen octahedra rotate out-of-phase about the $c$-axis; the imaginary phonon at $X$ point is an anti-polar mode with Co and O atoms forming an ``out-of-phase local polarization'' that alternates its direction unit cell by unit cell along $a$ axis. We note that due to the $C_4$ rotation symmetry of the simple tetragonal structure, $Y$ point is equivalent to $X$ point in the Brillouin zone and there is another imaginary anti-polar mode at $Y$ point with Co and O atoms forming an ``out-of-phase local polarization'' that alternates its direction unit cell by unit cell along $b$ axis. Introducing those phonon modes into the simple tetragonal structure will lower the total energy and yield a new crystal structure, in which a ferromagnetic polar metallic state may be stabilized.

\begin{figure}[t]
\centering
\includegraphics[width=0.8\textwidth]{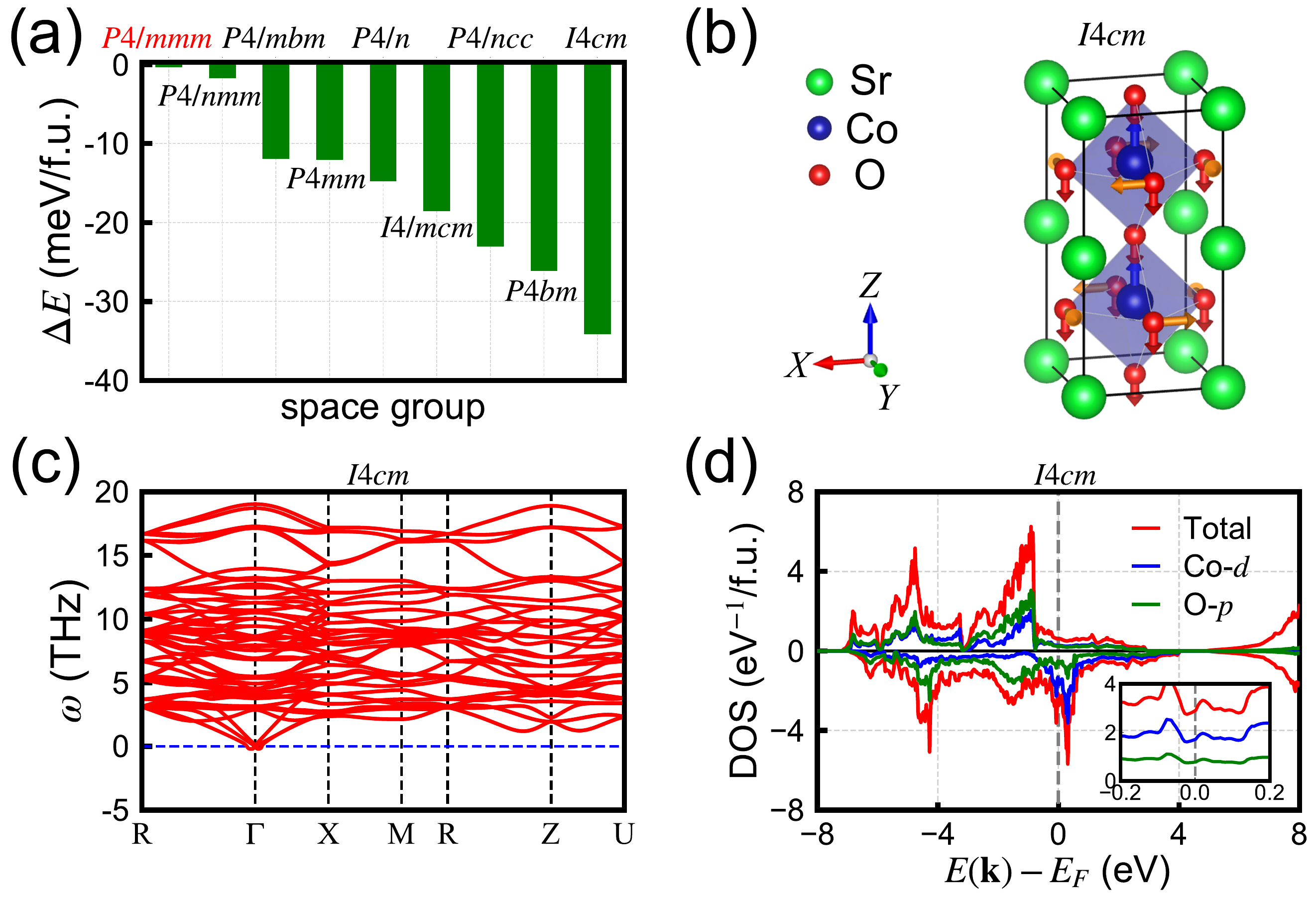}
\caption{(a): Total energy of different crystal structures of SrCoO$_3$ under 4\% compressive strain, using the simple tetragonal structure $P4/mmm$ as the zero point. (b): The crystal structure of $I4cm$ SrCoO$_3$. The arrows show that the structure has both polar displacements along the $c$-axis and an out-of-phase oxygen octahedral rotation about the $c$-axis. (c): Phonon band structure of $I4cm$ SrCoO$_3$. (d): Density of states (DOS) of $I4cm$ SrCoO$_3$. The red, blue and green curves are total, Co-$d$ projected and O-$p$ projected DOS. Inset: near-Fermi-level DOS (spin up + spin down) of $I4cm$ SrCoO$_3$.}
\label{fig3}
\end{figure}

To find the most stable crystal structure of SrCoO$_3$ that arises from the above imaginary phonon modes, we study 4\% compressive strain and introduce each imaginary phonon mode as well as their combinations into the simple tetragonal structure (a similar analysis is done on 3\% compressive strain and see Supplemental Material~\cite{SI} Note 4). 
Since the imaginary phonons at \textit{X} and \textit{Y} points are degenerate, we combine them together and consider it as a ``composite'' phonon, referred to as \textit{X/Y}. Thus we have 4 imaginary phonons at $\Gamma$, \textit{X/Y}, \textit{M} and \textit{R}. Their combinations (including one phonon mode) yield altogether $C_4^1 + C_4^2 + C_4^3 + C_4^4 = 15$ different cases. However, from our calculations, we find some imaginary phonon modes suppress each other, i.e. when two such imaginary phonons are combined and introduced into the high-symmetry structure, after structural relaxation we end up with a low-symmetry structure that is identical to the one that is derived only from one imaginary phonon. Specifically, we find that the imaginary phonon at $R$ point suppress the imaginary phonon at $M$ point, and the imaginary phonon at $\Gamma$-point suppresses the imaginary phonon at $X/Y$-point. Excluding those cases, we finally end up with 8 low-symmetry structures by introducing the imaginary phonons and their combinations. We explicitly list below all the 8 low-symmetry structures as well as the associated imaginary phonons:

\begin{enumerate}
\item The first low-symmetry structure is obtained by introducing the $\Gamma$-point imaginary phonon mode. It is a polar structure. The corresponding space group is $P4mm$ (no. 99) and Glazer notation is $a^0a^0c^0$.
  
\item The second low-symmetry structure is obtained by introducing the $M$-point imaginary phonon mode. It is a centrosymmetric structure with an in-phase rotation of CoO$_6$ oxygen octahedra about the $c$-axis. The corresponding space group is $P4/mbm$ (no. 127) and Glazer notation $a^0a^0c^+$. 

\item The third low-symmetry structure is obtained by introducing the $R$-point imaginary phonon mode. It is a centrosymmetric structure with an out-of-phase rotation of CoO$_6$ oxygen octahedra about the $c$-axis. The corresponding space group is $I4/mcm$ (no. 140) and Glazer notation $a^0a^0c^-$. 

\item The fourth low-symmetry structure is obtained by introducing $X$/$Y$-point imaginary phonon modes into the simple tetragonal structures. It is an anti-polar structure. The corresponding space group is $P4/nmm$ (no. 129) and Glazer notation is $a^0a^0c^0$.
  
\item The fifth low-symmetry structure is obtained by introducing $M$-point and $X/Y$-point imaginary phonon modes. It is a complicated anti-polar structure with an in-phase rotation of CoO$_6$ oxygen octahedra about the $c$-axis. The corresponding space group is $P4/n$ (no. 85) and Glazer notation is $a^0a^0c^+$.

\item The sixth low-symmetry structure is obtained by introducing $R$-point and $X/Y$-point imaginary phonon modes. It is a complicated anti-polar structure with an out-of-phase rotation of CoO$_6$ oxygen octahedra about the $c$-axis. The corresponding space group is $P4/ncc$ (no. 130) and Glazer notation is $a^0a^0c^-$. 

\item The seventh low-symmetry structure is obtained by introducing $\Gamma$-point and $M$-point imaginary phonon modes. It is a polar structure with an in-phase rotation about the $c$-axis. The corresponding space group is $P4bm$ (no. 100) and Glazer notation $a^0a^0c^+$.

\item The last low-symmetry structure is obtained by introducing $\Gamma$-point and $R$-point imaginary phonon modes. It is a polar structure with an out-of-phase rotation about the $c$-axis. The corresponding space group is $I4cm$ (no. 108) and Glazer notation $a^0a^0c^-$.
  
\end{enumerate}

Fig.~\ref{fig3}(a) shows the total energy of those new crystal structures (using the simple tetragonal structure $P4/mmm$ as the zero point). We find that the crystal structure of the lowest total energy is the complicated polar structure $I4cm$, which is explicitly shown in Fig.~\ref{fig3}(b). To further check that the $I4cm$ structure is indeed dynamically stable, we perform a phonon calculation on the $I4cm$ structure and find no imaginary modes in the phonon spectrum of $I4cm$ SrCoO$_3$, as shown in Fig.~\ref{fig3}(c). In Fig.~\ref{fig3}(d), we show the DOS of $I4cm$ SrCoO$_3$. There is a clear exchange splitting and a finite DOS at the Fermi level. Combining the electronic, magnetic and structural properties shown in Fig.~\ref{fig3}, we find that under 4\% compressive strain, a ferromagnetic polar metallic state is stabilized in $I4cm$ SrCoO$_3$. In addition, as we show below, there is a finite range of compressive strain in which SrCoO$_3$ exhibits a ferromagnetic polar metallic state.

\subsection{Tensile strain}

After studying SrCoO$_3$ under compressive strain, now we switch to tensile strain. We also find a ferromagnetic polar metallic state in SrCoO$_3$ when the applied tensile strain is appropriate. However, there are some important differences in the nature of polarity.

\begin{figure}[t]
\centering
\includegraphics[width=0.8\textwidth]{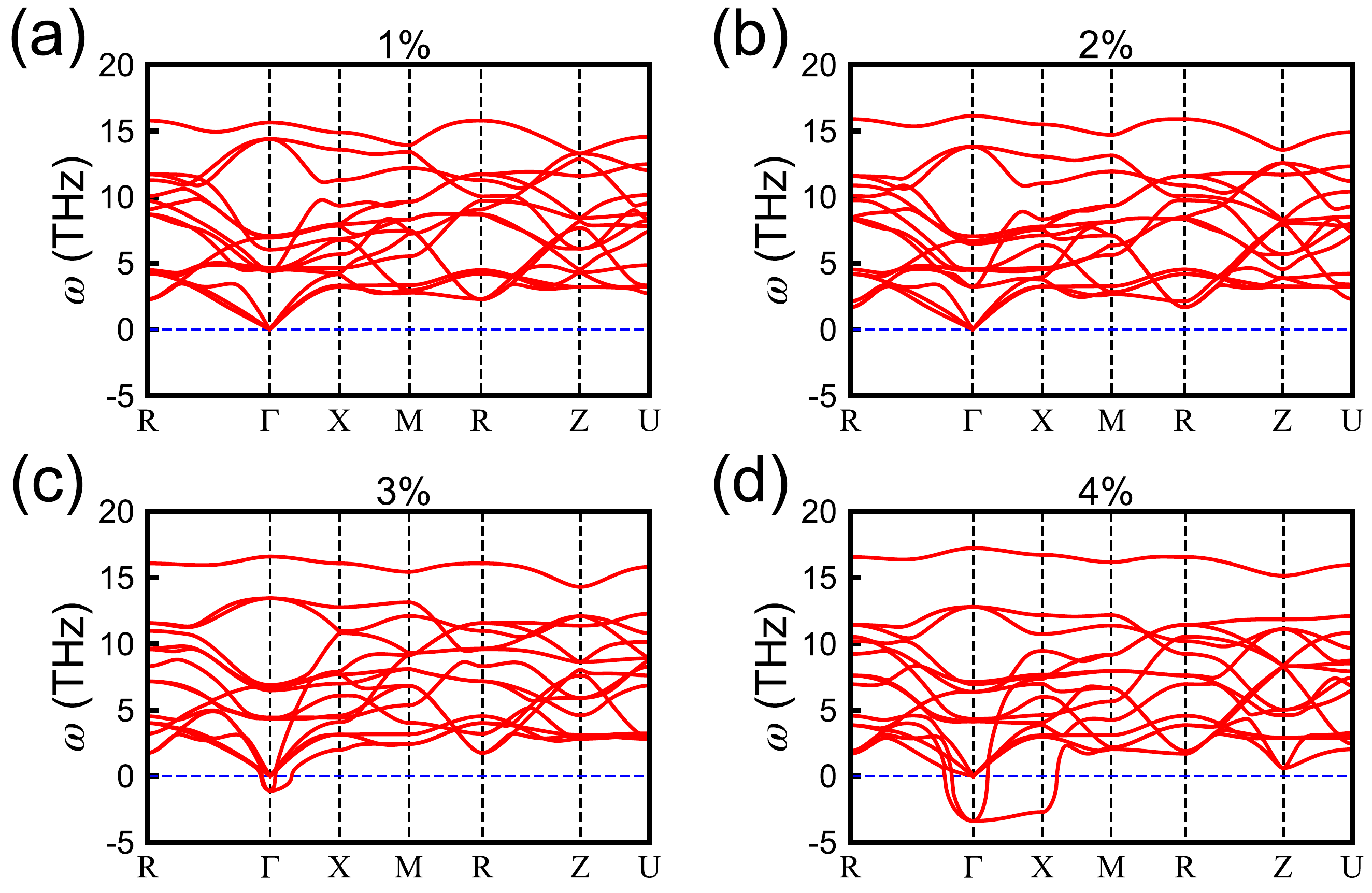}
\caption{Phonon band structure of SrCoO$_3$ under different tensile strains. (a): 1$\%$ tensile strain. (b): 2$\%$ tensile strain. (c): 3$\%$ tensile strain. (d): 4$\%$ tensile strain. The coordinates of the high-symmetry \textbf{k}-points are: $R$(0.5, 0.5, 0.5), $\Gamma$(0.0, 0.0, 0.0), $X$(0.5, 0.0, 0.0), $M$(0.5, 0.5, 0.0), $Z$(0.0, 0.0, 0.5), $U$(0.5, 0.0, 0.5).}
\label{fig4}
\end{figure}

Similar to compressive strain, SrCoO$_3$ under tensile strain also naturally transforms to a simple tetragonal structure but with $c/a < 1$. We calculate the phonon spectrum of SrCoO$_3$ in the simple tetragonal structure under various tensile strains. The results are shown in Fig.~\ref{fig4}. We find that under 1\% and 2\% tensile strains, the phonon spectrum of SrCoO$_3$ is free from imaginary phonon modes. However, under 3\% and 4\% tensile strains, imaginary phonon modes appear at $\Gamma$ and $X$ points. By analyzing the vibration modes, we find that the imaginary phonons at $\Gamma$ point are two-fold degenerate. They are both polar modes that are associated with the ``polarization'' of Co and O atoms along $a$ and $b$ axes, respectively. The imaginary phonon at $X$ point is an anti-polar mode with Co and O atoms forming an ``out-of-phase local polarization'' that points parallel to $b$-axis and alternates its direction unit cell by unit cell along $a$ axis. We note that due to the $C_4$ rotation symmetry of the simple tetragonal structure, $Y$ point is equivalent to $X$ point in the Brillouin zone and there is another imaginary anti-polar mode at $Y$ point with Co and O atoms forming an ``out-of-phase local polarization'' that points parallel to $a$-axis and alternates its direction unit cell by unit cell along $b$ axis.

Then we study 4\% tensile strain and introduce each imaginary phonon mode and their combinations into the simple tetragonal structure (a similar analysis is done on 3\% tensile strain and see Supplemental Material~\cite{SI} Note 4). Again since the imaginary phonons at $X$ and $Y$ points are degenerate, we combine them together and consider it as a ``composite'' phonon, referred to as \textit{X/Y}. Similar to the case of 4\% compressive strain, we find that the phonon vibration mode at $\Gamma$ point suppresses the phonon vibration mode at $X/Y$-point. Therefore, altogether we find two new structures whose energy is lower than that of the simple tetragonal structure.

\begin{enumerate}
 \item One is a polar structure with no oxygen octahedral rotations, by introducing the $\Gamma$-point polar mode into the simple tetragonal structure. The corresponding space group is $Amm2$ (no. 38) and Glazer notation is $a^0a^0c^0$. 
 \item The other is an anti-polar structure with no oxygen octahedral rotations either, by introducing both $X$-point and $Y$-point anti-polar modes into the simple tetragonal structures. The corresponding space group is $P4/mbm$ (no. 127) and Glazer notation is $a^0a^0c^0$.
\end{enumerate}

\begin{figure}[t]
\centering
\includegraphics[width=0.8\textwidth]{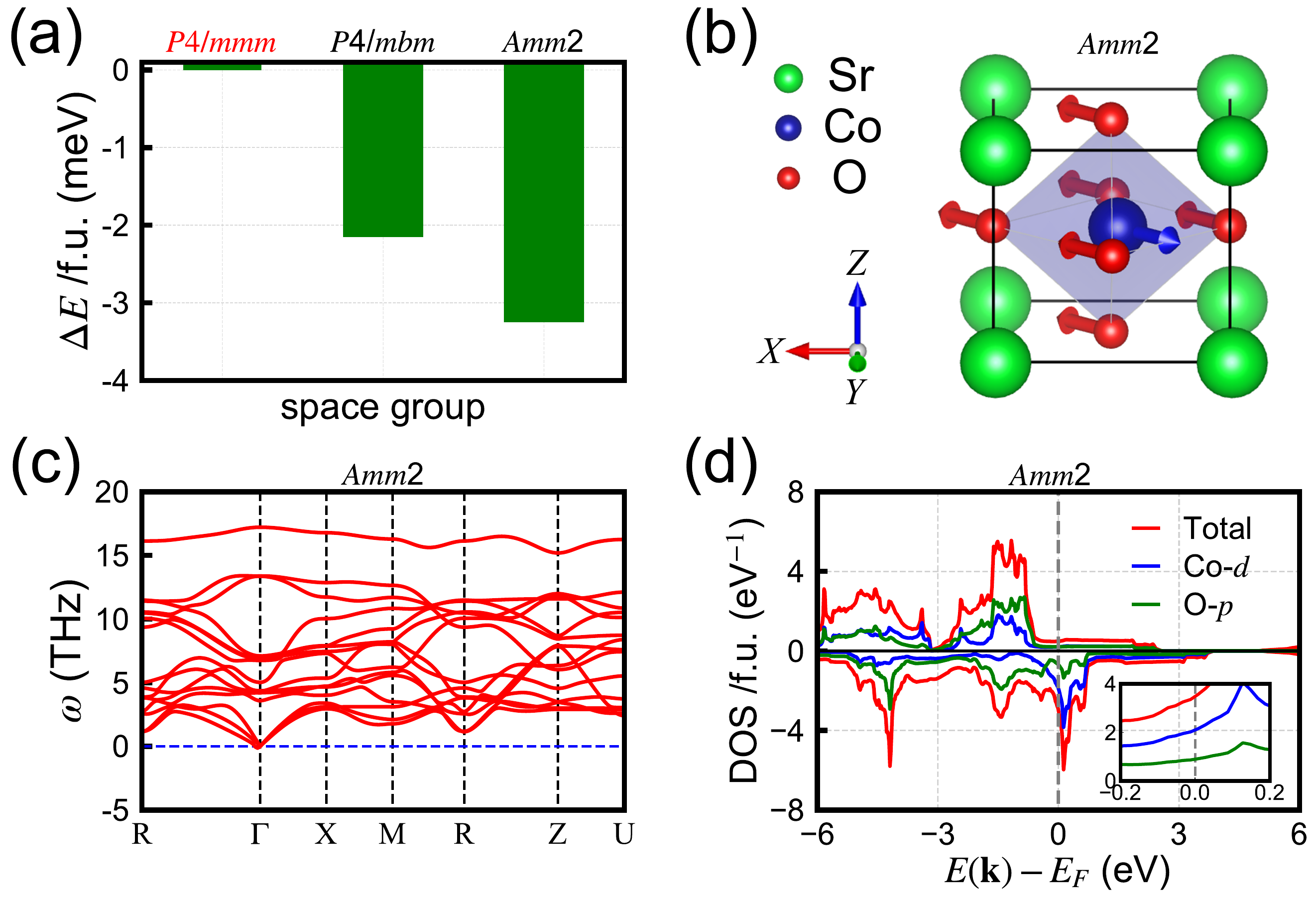}
\caption{(a): Total energy of different crystal structures of SrCoO$_3$ under 4\% tensile strain, using the simple tetragonal structure $P4/mmm$ as the zero point. (b): The crystal structure of $Amm2$ SrCoO$_3$. The arrows show that the structure has polar displacements along the diagonal line in the $ab$ plane. (c): Phonon band structure of $Amm2$ SrCoO$_3$. (d): Density of states (DOS) of $Amm2$ SrCoO$_3$. The red, blue and green curves are total, Co-$d$ projected and O-$p$ projected DOS. Inset: near-Fermi-level DOS (spin up + spin down) of $Amm2$ SrCoO$_3$.}
\label{fig5}
\end{figure}

Fig.~\ref{fig5}(a) shows the total energy of these two new crystal structures, using the simple tetragonal $P4/mmm$ as the zero point. We find that the $Amm2$ structure has the lowest total energy. Fig.~\ref{fig5}(b) explicitly shows the $Amm2$ structure in which the ``polarization'' lies along the diagonal line of $ab$ plane. We also test whether this new $Amm2$ structure is dynamically stable. Fig.~\ref{fig5}(c) shows the phonon spectrum of $Amm2$ SrCoO$_3$ and we find no imaginary phonon modes. Fig.~\ref{fig5}(d) shows the DOS of $Amm2$ SrCoO$_3$, which has a clear exchange splitting and a finite value at the Fermi level. Similar to $I4cm$ SrCoO$_3$ under 4\% compressive strain, combining the electronic, magnetic and structural properties shown in Fig.~\ref{fig5}, we find that under 4\% tensile strain, a ferromagnetic polar metallic state is also stabilized in $Amm2$ SrCoO$_3$. As we show below, there is also a finite range of tensile strain in which SrCoO$_3$ exhibits a ferromagnetic polar metallic state.

\begin{figure}[t]
    \centering
    \includegraphics[width=0.8\textwidth]{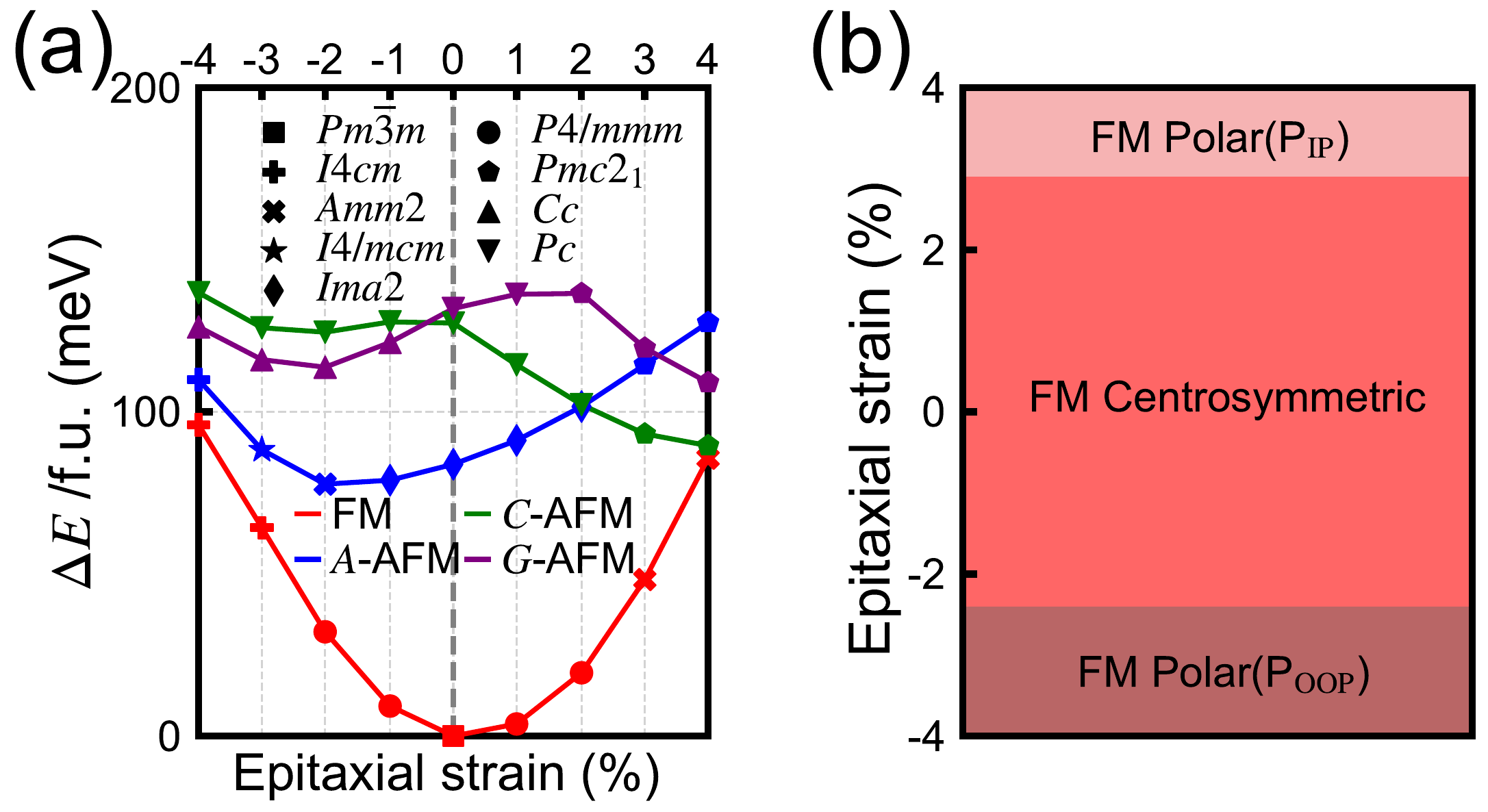}
    \caption{(a): An energy diagram of SrCoO$_3$ as a function of epitaxial strain. Red, blue, green and purple colors represent ferromagnetic order (FM), $A$-type antiferromagnetic order ($A$-AFM), $C$-type antiferromagnetic order ($C$-AFM) and $G$-type antiferromagnetic order ($G$-AFM). Different symbols represent different crystal structure symmetries (the space group is shown). The energy of cubic FM SrCoO$_3$ is used as the zero point. (b): A phase diagram of SrCoO$_3$ as a function of epitaxial strain. ``P$_{\rm{IP}}$'' means ``polarization'' lying in the $ab$-plane and ``P$_{\rm{OOP}}$'' means ``polarization'' pointing along the $c$-axis.}
    \label{fig6}
\end{figure}

\subsection{Magnetic transition}

%Finally, we discuss magnetic transition of SrCoO$_3$ under different epitaxial strains.

In the preceding calculations, we assume that SrCoO$_3$ is in a ferromagnetic metallic state. Previous studies have shown that under biaxial strain, SrCoO$_3$ exhibits a ferromagnetic-to-antiferromagnetic transition as well as a metal-insulator transition~\cite{10.1103/PhysRevLett.107.067601,10.1103/PhysRevB.91.140405,10.1103/PhysRevX.10.021030}. Following Ref.~\cite{10.1103/PhysRevLett.107.067601}, we consider three common types of antiferromagnetic ordering: $A$-type with an ordering wave vector $(0,0,\pi)$, $C$-type with an ordering wave vector $(\pi,\pi,0)$ and $G$-type with an ordering wave vector $(\pi,\pi,\pi)$). For all these magnetic orderings, we consider various structural distortions with different orientations of ``polarization'' and different types of oxygen octahedral rotations. After atomic relaxation, for each type of magnetic ordering under a given epitaxial strain, we obtain the most stable crystal structure. Then we compare the total energies of those crystal structures with different types of magnetic ordering. For all the strains considered in this study, we find that SrCoO$_3$ remains metallic. We show the results in Fig.~\ref{fig6}(a). We find that no matter whether SrCoO$_3$ is in a ferromagnetic state or in an antiferromagnetic state, its most stable structure undergoes a series of distortions and changes in crystal symmetry with an applied biaxial strain. More importantly, within 4\% compressive or tensile strain, the ferromagnetic ordering always has lower energy than the antiferromagnetic orderings. Fig.~\ref{fig6}(b) summarizes the phase diagram of SrCoO$_3$ as a function of epitaxial biaxial strain. Under a compressive strain from 2.4\% to 4\%, SrCoO$_3$ is in a ferromagnetic polar metallic state with ``polarization'' along the $c$-axis (dark red range). Under a tensile strain from 2.9\% to 4\%, SrCoO$_3$ is also in a ferromagnetic polar metallic state with ``polarization'' lying in the $ab$-plane (light red range). In between, SrCoO$_3$ is a ferromagnetic metal with inversion symmetry. Finally, we note that if the applied strain is larger than 4\%, a ferromagnetic-to-antiferromagnetic transition may occur to SrCoO$_3$ (see Supplemental Material~\cite{SI} Note 5 for details).

%However, when SrCoO$_3$ under 5$\%$ tensile strain, the crystal structure of SrCoO$_3$ have in-plane polarization with C type antiferromagnetic ordering (green range). Therefore, when the compressive strain of SrCoO$_3$ is large enough, the crystal structure will appear out-of-plane polarization. However, when the tensile strain is large enough, the crystal structure will appear in-plane polarization. In our calculations, the structures of  SrCoO$_3$ under 5$\%$ tensile strains exhibit in-plane polarization with 0.341 $\mathrm{\AA}$ displacement of Co$^{4+}$ ions and oxygen octahedral rotation ($a^-a^-c^+$). In order to prove the structural stability of SrCoO$_3$ under antiferromagnetic ordering, we further calculate the phonon band structure. Panel \textbf{c} shows the phonon band structure of C-AFM SrCoO$_3$ under 5$\%$ tensile strain. We can see that there are no negative phonon frequency on all high-symmetry \textbf{k}-points. This indicates that the structure of C-AFM SrCoO$_3$ is stable under 5$\%$ tensile strain. And it maintain metallic properties(shown in panel \textbf{d}). Therefore, SrCoO$_3$ transform into antiferromagnetic polar metals. In addition, When the tensile strain is larger, The SrCoO$_3$ will transform from C-AFM to G-AFM (see the Supplementary section 3).

\section{Conclusion}

In summary, via first-principles calculations, we show that epitaxial strain engineering provides a simple alternative to synthesizing new materials in the search for magnetic polar metals. We demonstrate this route in a ferromagnetic metallic oxide SrCoO$_3$. We find that using either a compressive or tensile strain of experimentally feasible magnitude can induce a ferromagnetic polar metallic state in SrCoO$_3$. Specifically, under a compressive strain of 2.4\%-4\%, an $I4cm$ structure is stabilized in SrCoO$_3$ with the ``polarization'' pointing along the $c$-axis as well as an out-of-phase oxygen octahedral rotation about the $c$-axis. Under a tensile strain of 2.9\%-4\%, an $Amm2$ structure is stabilized in SrCoO$_3$ with the ``polarization'' lying in the $ab$-plane but with no oxygen octahedral rotations. In addition, we find that under sufficiently large epitaxial strain ($> 4\%$), a ferromagnetic-to-antiferromagnetic transition may occur to SrCoO$_3$. Such a large strain is challenging via epitaxy but might be achieved in freestanding thin films~\cite{10.1126/science.aax9753}. We hope that our work may stimulate further theory and experiment studies on the search for magnetic polar metals. 

\begin{acknowledgments}
  We are grateful to Pu Yu for useful discussions. This project was
  financially supported by the National Key R\&D Program of China
  under project number 2021YFE0107900, the National Natural Science
  Foundation of China under project number 12374064, Science and Technology
  Commission of Shanghai Municipality under grant number 23ZR1445400
  and a grant from the New York University Research Catalyst Prize.
  NYU High-Performance-Computing (HPC) provides computational
  resources.
\end{acknowledgments}

\newpage
\clearpage

%\bibliographystyle{naturemag}

%\bibliographystyle{apsrev}
%\bibliography{reference-3}

\end{document}